\documentstyle[11pt,aaspp]{article}
\input epsf
\def\l*{$L_*$\/}
\def\etal{{\it et al. }}
\def\vcirc{{v_{circ}}}
\def\kms{\rm km~s^{-1}}

\def\lsim{\mathrel{\lower .85ex\hbox{\rlap{$\sim$}\raise
.95ex\hbox{$<$} }}}
\def\gsim{\mathrel{\lower .80ex\hbox{\rlap{$\sim$}\raise
.90ex\hbox{$>$} }}}

\begin{document}

\title{The Formation of Quasars in Low Luminosity Hosts via
Galaxy Harassment}

\vskip 0.5truecm

\centerline{\bf George Lake$^{\bf 1}$, Neal Katz$^{\bf 2}$ and Ben Moore$^{\bf 3}$}

{\bf 1.} {Astronomy Department, University of Washington, Seattle, 
WA 98195, USA}

{\bf 2.} {Department of Physics \& Astronomy, University of Massachusetts, 
Amherst, USA}

{\bf 3.} {Department of Physics, University of Durham, Durham, UK}

\begin{abstract}
   
We have simulated disk galaxies
undergoing 
continual bombardment by other galaxies in 
a rich cluster.  ``Galaxy harassment" leads to    
dramatic evolution of smaller disk galaxies  
and provides an extremely effective 
mechanism to fuel a central quasar.
Within a few billion years after a small
disk galaxy enters the cluster environment,  
up to 90\% of its gas can be driven into 
the inner 500 pc.   Up to 
half of the mass can be transferred in a burst 
lasting just 100-200 Myr.
This transport of gas to the center of
galaxy is far more efficient than any mechanism proposed before.
Galaxy harassment was first proposed to explain the disturbed blue
galaxies in clusters seen in clusters 
at ($z \gsim 0.3$),  the  ``Butcher--Oemler effect".
Quasars at the same reshifts 
lie in more clustered
environments than those at lower redshift. 
Recent HST observations find that roughly 
half of all observed quasar host galaxiess are fainter
than \l*, with many of these less luminous hosts occuring
at redshifts $z \gsim 0.3$.
We examine 5 quasars that are claimed to have 
low luminosity hosts 
and find that 3 are in rich clusters of galaxies, the fourth
may be in a cluster but the evidence for this is marginal. The environment
of the fifth has not been studied.

\end{abstract}


\section{Introduction}

To power the central black hole by accretion, 
bright quasars must consume $10^8-10^{9}$ solar masses during their
lifetimes---roughly 1\% of the stellar mass in a bright elliptical
galaxy or 10\% of the gaseous mass of a bright spiral.
Even with the assumption of a large host galaxy, an
efficient mechanism is needed to 
channel gas to the central source.   
This problem of fuelling quasars is normally divided into three parts:
the movement of gas from galactic scales to the inner
few hundred parsecs, the instabilities of a 
self gravitating disk that
transports the gas to a compact accretion disk
and the detailed dynamics and radiation 
mechanisms of the final accretion.
Here, we consider the first 
problem of moving gas from galactic scales into the
inner few hundred parsecs.
At redshifts of 2 where the number densities of quasars peak,
this has been associated with the dynamical chaos of galaxy
formation (c.f. Haehelt and Rees 1993).  
At lower redshifts, interactions of galaxies leading
to coalescence or merging have been proposed (Hernquist 1989).
These mergers can drive more than 10\% of the gas to the center.
The final host galaxy is the product of a merger, therefore it
would be brighter than average.

There are two observations that suggest a third mechanism at
intermediate redshifts $0.2 \lsim z \lsim 0.8$.
The environment 
of quasars has been observed to change 
with redshift (Yates,  Miller and Peacock 1989; hereafter
YMP89, 
Yee and Ellingson 1993; hereafter YE93).
Quasars at higher redshifts 
are in Abell richness class 0-1 
clusters of galaxies---an environment that is considerably richer
than that of 
lower redshift quasars  (the break point is at $z \sim 0.3$ in YMP89
and $z \sim 0.6$ in YE93).  
Bahcall, Kirhakos and Schneider (1995; hereafter
BKS) used HST to image 
eight luminous quasars at redshifts between 0.15 and
0.3.  Only three quasars in their sample 
have candidate hosts that are
as luminous as {\l*}, the characteristic luminosity in the Schechter (1976)
luminosity function (LF).  The other five 
hosts must be fainter than  {\l*} to have escaped detection. 
 
After observations of additional 
quasars,  Bahcall {et al.} (1997; hereafter BKSS97) 
conclude that ``the luminous quasars 
studied in this paper occur preferentially in luminous galaxies".
They are rejecting the ``null hypothesis" that all galaxies are equally
likely to have quasars (e.g. a hypothesis that states that Draco
and the Milky Way are equally likely to host quasars).
Their conclusion results because at least half of all galaxies are
$\gsim 2$ magnitudes fainter than \l* whereas the dividing line for 
their sample of quasar hosts is $\sim${\l*} within their errors.
This depends slightly on the logarithmic slope of the LF.
Since most LFs are weakly diverging at the faint end $N \propto
L^{-x}$, $1.5>x>1$, a faint end cutoff is required to
to define an ``average luminosity" and that average is normally
2 to 3 magnitudes brighter than the cutoff or 
$\gsim 2$ magnitudes fainter than {\l*}.
  
However, galaxies brighter than $0.7-0.8${\l*}
contain half of all the luminosity, where the range
takes into accounts the uncertainty in the faint end slope
and cutoff of the luminosity function.  This dividing line of luminosity
is consistent with the BKSS97 midpoint of quasar hosts within their errors.
So, quasars don't obviously prefer brighter galaxies any more than 
stars do.  Previously, McLeaod and Rieke (1995) suggested a linear
relation between the absolute magnitude of the quasar and its host.
BKSS97 find no such relation other than what can be attributed
to the obvious bias from detection limits.   
The simplest summary of the observations to date
is that quasars and galaxies may be related much as stars and galaxies; the
probability of finding either in a galaxy is proportional to the galaxies
luminosity but their individual luminosities are not determined by the
luminosity of their host.  Hence, {\it we need a mechanism that will
produce quasars with a frequency per unit luminosity rather than a mechanism
that only operates in bright galaxies as would be the case if mergers
were the dominant trigger}.

We have found a mechanism  that is  extremely efficient at channeling
gas into the center of sub-\l* galaxies that live in clusters.
``Galaxy harassment" 
drives dynamical instabilities that send most of the gas into the
central few hundred parsecs of the harassed galaxies.  
In \S 2, we present detailed hydrodynamical
simulations that show these effects, while we
look at the model's predictions in \S 3.

\section{Description of the Simulations}

We use TREESPH (Hernquist and Katz 1989, Katz, Weinberg
and Hernquist 1996) to examine the fate of a high resolution 
galaxy with $2^{14}$ particles in each of three components:  
gas, stars and dark matter.
The simulations were originally performed to follow 
the morphological evolution of galaxies in clusters (Moore \etal 1996, 
Moore, Katz and Lake 1996b),
so the parameters were chosen to be typical of the disturbed
galaxies seen by HST at redshifts of $\sim 0.3$. 
We present three simulations, two
with circular velocities of 
$160 \kms$ and one with a circular velocity of $110 \kms$.  
Using the Fisher--Tully (1981) relation,
the luminosity of a galaxy with a circular velocity, $\vcirc = 160 \kms$
is  $\sim L_*/5$.   
The $L_*/5$ model galaxies  have
exponential disks with scalelengths of 
2.5 kpc and
scaleheights of 200 pc.  The smaller galaxy is scaled by using
the Faber-Jackson relation and preserving the ratio of stars, gas
and dark matter.
The disks are constructed with a Toomre (1964)
``stability'' parameter $Q=1.5$ and  run in isolation for 2 Gyr
before being set into orbit in the cluster.  The gaseous disk
is initially on cold circular orbits.  

The dark halo of each galaxy is a spherical
isothermal with a core radius of $(v_{circ}/160\kms)^2$ kpc,
and is tidally truncated at
the pericenter of the galaxy's orbit within the cluster.  
At 4 disk scalelengths (10 kpc), the initial ratio of dark matter to stars to gas
is 11:4:1.   At 8 disk scalelengths (20 kpc), the ratios are
20:5:1 and the total mass is $\sim
10^{11} (v_{circ}/160\kms)^3 M_\odot$.

The cluster model is based on Coma.  Its
one dimensional velocity dispersion is 
$\sigma_c=1,000 \kms$ and the
mass within the virial radius (1.5 Mpc) is $7\times
10^{14}M_\odot$.  For an $M/L$ of 250, the total
cluster luminosity within this radius is $2.8\times 10^{12}L_\odot$.
The other galaxies in the cluster are drawn from 
a Schechter (1976) luminosity function parameterized using
$\alpha=-1.25$ and $M_*=-19.7$  
including all galaxies brighter than
$2.8\times 10^8L_\odot$ ($H_o=100{\kms Mpc^{-1}}$ and $\Omega=1$
throughout this paper). 
This produces a  model cluster that has 950
galaxies brighter than the Magellanic clouds, but only 31 brighter
than $L_*$.   

The masses and tidal radii of the other galaxies are determined
by taking an
isothermal model with a dispersion given by the
Faber-Jackson relation and 
tidally limiting it at the galaxy's pericentric distance.
They are then modeled by spheres with a softening length
equal to half of their tidal radius.  Most simulations have
no ``interpenetrating" collisions.  However, any such collision
will be more gentle than a collision with a realistic model
of a galaxy that is more centrally concentrated.  
White and Rees (1978) speculated that the dark halos were stripped
from galaxies within clusters;  we find that the dominant
stripping mechanisms are tides and high speed 
encounters---``galaxy harassment" (Moore, Katz and Lake 1996b;  hereafter MKL96b).  
If galaxies are initially 
tidally limited by the cluster potential, 
bright galaxies retain more
than half of their mass when followed for 5 Gyr.  
For greater self-consistency,  we reduced the mass of each
perturber by 25\%, the average loss over a Hubble time (MKL96b) and
we left the tidal/softening radius fixed.  In the accompanying
video, the perturbing galaxies are shown as green dots located
at their centers.

The mean ratio of a perturbing
galaxy's apocenter to its pericenter in our cluster model is roughly 6-to-1
(this ratio is even larger in infinite isothermal spheres with isotropic
velocities);   a galaxy found at a radius of 450 kpc will have a 
mean orbital radius of 400 kpc and a 
typical pericenter ($r_{peri}$) that is slightly greater than 150 kpc.  
We assign galaxies masses of $2.8\times10^{11} (r_{peri}/150{\rm kpc})
(L/L_*)^{3/4} M_\odot$ corresponding to mass-to-light ratios, 
$M/L = 26 h^2  (r_{peri}/150{\rm kpc})(L/L_*)^{-1/4}$.
The luminous parts of elliptical galaxies  have mass-to-light ratios 
of $12  h M_\odot/L_\odot$ (van der Marel 1991), so our 
perturbing galaxies have very modest
extended dark halos. 

The fraction of the cluster's density attached to
galaxies varies with radius from zero at the center to nearly
unity at the virial radius.
It is  $\sim$20\% at the mean orbital radius of our simulated
galaxies.  
The rest of the cluster mass is in a smoothly
distributed background represented by a fixed analytic potential.
Further details of the cluster model 
can be found in MKL96b.

We have intentionally made some conservative assumptions 
to ensure that our results are robust.  
For a model galaxy on a fixed orbit, 
the havoc wreaked by harassment depends on the square of the masses
of the largest galaxies encountered.
The most massive galaxies are  giant ellipticals
that are far less prone to harassment owing to their high 
internal densities,
yet we have reduced the masses of all galaxies by the same 
time averaged value of 25\%.  
At a fixed mean orbital radius, 
galaxies on elongated orbits experience greater harassment.  
We follow galaxies that have apo/peri ratios of 2 ({\it i.e.} apocenter
at 600 kpc, pericenter at 300 kpc), whereas the typical value is
$\sim 6$.  
As a result, our model galaxies avoid
extremes of the cluster distribution and start with large dark halo
masses determined by the tidal limit at their atypically large pericenters.  
Both effects serve to underestimate the effects of
harassment.

The strongest encounters do not necessarily
occur near the center of the cluster.  Clearly, the frequency of encounters
scales with the density of galaxies.  However, there are two
counterbalancing effects.
If  the galaxies are all tidally limited, they are more massive
at larger cluster radii.    
Secondly, the relative velocity of encounters decreases in
the outer parts of the cluster and 
encounters are stronger at fixed
impact parameter.  
As long as the encounter timscale is short compared to the
internal dynamical time of the galaxy, 
the gas dynamical effects are  
strongest outside the central region of a cluster.   
However, galaxies near the center of the cluster
are more strongly damaged by the cluster's tidal field.

Care was taken to ensure that our initial models were sensible
and stable by simulating galaxies on
circular and elliptical orbits in smooth cluster potentials 
before examining the effects of harassment
by other galaxies.
The disk of a galaxy on circular orbit at 450 kpc in a fixed
cluster potential remains
stable for 10 Gyrs.  A galaxy on an eccentric orbit with
an apocenter of 600 kpc and pericenter of 300 kpc  becomes
bar unstable after the first pericentric passage and a bar 
persists for 5 billion years.  Each
passage through pericenter results in the loss of a small fraction
of dark halo material, but the gas and stars remain attached.

The most dramatic evolution owes to strong encounters
when the other perturbing galaxies are included
(one simulation is shown on an accompanying video).   
The first strong encounter 
usually causes a bar instability of greater strength than
that induced by only the cluster's tidal field.
The continued heating of the disk by collisions 
transforms the galaxy into  a spheroidal galaxy that
matches the 
brightness profile and velocity dispersions of the
dwarfs in nearby clusters (Moore, Lake and Katz 1997).  

Perhaps the most stunning feature of the evolution is the rate that
gas is driven into the center of the galaxy, as shown in
Figures 1 and 2.  Figure 1 shows snapshots of the evolution of gas disk 
(also shown as the last sequence in the accompanying video).  The angular
momentum of the gas is decreased by the torques of
passing collisions and internal torques between the distorted
disk and the halo.  When combined with radiative cooling, this drives a large
fraction of the gas into the center of the galaxy as shown in 
Figure 2.   In our first three
simulations, we found that $\sim$90\%,  80\% and 40\%
of the gas is driven to the center within 3 Gyrs.  In the first two cases,
half of the mass is transferred in an interval of 100-200 Myr.

We note one last problem in making detailed comparisons to observations.
We link the collisional deformation of a galaxy to the channeling of
gas into the middle of the galaxy.  
The response to the jolt of a high speed
fly-by encounter occurs over the galaxy's 
internal orbital time, $\gsim 200$ Myr
(for a half light radius of 5.6 kpc and circular velocity of $160 \kms$).   
The three dimensional velocity dispersion in the cluster
is $\sim 1,700 \kms$, so each galaxy moves 
$\sim 400$ kpc through the cluster
as it responds to their encounter.
Therefore, one can determine neither the galaxy that stimulated the 
encounter nor the location where the encounter occurred.

We haven't included star formation in our gas dynamical 
simulations, a weakness that we share with all past models
of quasar fueling by gas in galaxies.
The lack of star formation  
enables the gas to radiate the random 
energy resulting from the impulsive torques that diminish its angular momentum.
If the material were all instantly turned into stars, it would
not continue its inward flow and the central
source would be extinguished.  This caveat applies
to all other models that use gas dyanmical simulations such as feuling with
mergers ({\it cf.} Hernquist).  However, we offer some observational
evidence that the material does remain gas for a fairly long time.

In an accompanying paper
(Moore, Lake and Katz 1997), we examine the remnants and compare
them to spheroidal galaxies.  
We will examine the possibility of ``black hole hunting" in
nucleated spheroidals in the next section.

\section{Predictions}

Galaxy harassment only occurs in clusters where the impact velocities
are too large to permit merging.  By invoking this process to explain
the qso hosts below \l*, we make four clear predictions that will
each be explored in the next sections:  
\begin{itemize}
\item{} quasars with low luminosity hosts should 
be in clusters where harassment occurs,  
\item{} resolved hosts should appear disturbed,
\item{} regions where harassment is ongoing should show an 
enhanced quasar frequency
\item{} black holes should exist in some nucleated spheroidal galaxies
\end{itemize}

\subsection{The Environment of the BKS95 sample}

Early studies of the environments of quasars concluded
that they were not in rich clusters (Roberts, O'Dell and Burbidge
1977, Stockton 1978, French and Gunn 1983).  That picture has changed
and it is now clear
that many low redshift quasars are in rich clusters.
The majority of these may lie
at the
edge of the cluster
(\cite{oem78}, Green and Yee 1984, Yee \etal 1989), but
there are a few such as H 1821+643 (Schneider \etal 1992)
and 3C 206 (Yee \etal 1989) that appear to lie at the center.
Many other quasars show
excesses of close optical companions (Hutchings, Crampton 
and Campbell 1984, Dahari 1984, Heckman \etal 1984 and
Yee 1986).  Heckman \etal (1984) obtained redshifts for 15 optical 
companions and found that 14 out of 15 were physical companions
not projections.  

Several groups have found that the clustering properties of 
quasars change rapidly with 
redshift.  YMP93
find that quasars with redshifts $z \gsim 0.3$ occur in 
Abell richness class 0-1 
clusters of galaxies, an environment that
is roughly three times richer than that of the
lower redshift quasars.   YE93 reach a similar
conclusion, but place the redshift break slightly higher ($z \gsim 0.6$).
Fisher \etal (1996; hereafter FBKS) looked at a sample of
($z \lsim 0.3$) with a mean redshift of $\sim 0.2$.  
They conclude  
that the quasars 
reside in stuctures that have at least half as many members as 
richness class R=0 Abell clusters.  So, there seems to be a smooth
trend with redshift.

These results agree qualitatively with our model,
where the chaos of galaxy formation remains the reason for the
observed peak of quasar number densities at a redshift of 2
and merging could dominate at the present day.  
In the standard hierarchical clustering model,
Kaufman (1995) has shown that the infall rate of field galaxies into
clusters peaks at redshifts of 0.3--0.5.
At these
redshifts, infalling galaxies are harassed to produce the central gas flow
needed to create a quasar.
The real test will be a measurement of the evolution of the
galaxy--galaxy and quasar--galaxy correlation functions to $z \sim 0.6$.

We can also examine what is known about the environments
of the quasars in the BKS95 sample.
BKS95 found 7 companions within 25 kpc of their targets, whereas they
estimate the chance occurrence should be 0.375 per target.  
We found a wealth of other information on the environments of 
these quasars. 
     
\noindent{\sl PG 0953+414:}  This is one of the richest environments
in the Green and Yee (1984) sample.  They see 25 galaxies in their
field of 140" (a radius of 1 Mpc centered on the quasar at $z=0.234$).  
BKS95 detect one companion, there are none
evident on the H-band image of McLeod and Rieke (1994, hereafter McR).

\noindent{\sl PG 1116+215:} This quasar lies at the edge of a 
relatively rich cluster
discovered by Green and Yee (1984a) and
studied by Ellingson, Green and  Yee (1991).  BKS95 found
one companion within 25 kpc.                                                                             
              
\noindent{\sl PG 1202+281:} BKS95 find two companions.  There is a 
bright companion that can be clearly seen on McR's H-band image.
We found no studies of it's broader environment.

\noindent{\sl 3C 273:} BKS95 detect a host with a brightness that is
greater than {\l*}, but they see no companions.
McR see the jet to NW in 
their H-band frame and a second feature that is just S of W.
Stockton (1980) found four galaxies within 250 kpc with velocity
differences of -80, 300, 530, and 510 $\kms$.  This quasar is
probably in a poor cluster.  

\noindent{\sl PKS 1302-102:} BKS95 found two companions within 25 kpc.
The limit set by BKS95 was just 0.4 magnitudes fainter than \l*.  This
quasar was also observed by Disney \etal (1995) who detected a host
galaxy using a different filter and detector (the Faint Object Camera
rather than the Wide Field/Planetary Camera).  When they apply a
standard color correction for the elliptical galaxy they detected, 
they find their detected galaxy is 0.2 magnitudes fainter than the
BKS95 limit.
This galaxy has been included in several studies of the environments
of quasars.  Green and Yee (1984) find 10 galaxies on their 140" field
frames.   Its cross-correlation with galaxy counts is typical of
an Abell richness class 0 cluster, but the significance of the 
cross-correlation is marginal (Yates,  Miller and Peacock 1989, 
Yee and Ellingson 1993).

\noindent{\sl PG 1307+085:} BKS95 did not detect any companions within
25 kpc, but Yee (1987) found one that is just 41 kpc away.  The H-band
image of McR shows two patches of luminosity that are offset from the position
of the quasar.

\noindent{\sl PG 1444+07:}  BKS95 detected a host galaxy with
a magnitude of {\l*}.  It appears to have a bar and ring.
The H-band image in McR shows luminosity that is offset from the QSO.
This is the only quasar in the BKS95 sample that was included in 
Green and Yee's (1984) sample, but no rich cluster was found.   

\noindent{\sl 3C323.1:} Oemler, Gunn and Oke (1972) 
declare that this is ``A QSO in a Rich Cluster of Galaxies".  It
lies 6.5 arcmin from the center of a compact Zwicky cluster.  This
is slightly less than 1 Abell radius.   BKS95 find a companion 6.9 kpc away
that was previously found by Stockton (1982) to have a velocity
difference with respect to the QSO of 
$\sim$150 $\kms$.  Three companions can be seen on McR's H-band image.

In summary, three of the quasars 
are known to live in 
rich clusters of galaxies: PG 0953+414, PG 1116+215, and 3C323.1.
A fourth (PKS 1302-102) appears to be in a cluster, but the evidence
is marginal.  
The only quasars that have deep wide field images and
do not appear to be in clusters have hosts that are
$\sim${\l*}.
The galaxy host of 3C273 is brighter than \l* and it is found in an
environment that might be more conducive to merging than harassment
(the velocity dispersion appears to be less than 500 $\kms$).
The only quasar in the Green and Yee (1984) survey
that does not appear to be in a cluster,  PG 1444+407, 
has the second most luminous host ($\sim${\l*}) in the BKS95 sample.

More recently, BKSS97  completed a study of 20 quasar
hosts.  In the original BKS95 sample, half of the quasars had reshifts between
0.239 and 0.236.  While the newer sample is 2.5 times larger, it
added only 2 more galaxies to the redshift range where we expect harassment
to become important. Examining the two new quasars in this range, 
one is observed
to be interacting (0316-416) and the other is an elliptical (PG1004+130).

\subsection{Characteristics of the Candidate Host Galaxies}

BKS95 only found candidate hosts for the 3 QSOs:  PG 1116+215, 3C 273 
and PG 1444+407.    Only the host of 
3C 273 is clearly as bright as {\l*} and it
appears to be an elliptical slightly offset from the position of 
the QSO.  The hosts of the other two galaxies are about a half magnitude
fainter than \l* when BKS95 use aperture magnitudes and slightly brighter
than \l* if they employ an exponential disk model.
The PG 1116+215 host appears to have partial rings while
the PG 1444+407 host appears to have a bar and a ring-like structure.
Having failed to confirm  past
detections of QSO hosts, BKS95 are circumspect in their descriptions of
their images.  However, we are encouraged by
the structures described for the two less
luminous hosts.  Strong bars and rings are often seen in harassed
galaxies.

\subsection{Are Quasars More Frequent Where Galaxies Are Being Harassed?}

The phenomena of galaxy harassment was discovered in our study
of the destruction of substructure in mergers (MKL96b).  
In Moore \etal (1996), we
first applied it to explain the evolution of cluster populations
discovered by Butcher and Oemler (1978).   
Harassment is clearly occurring in these clusters.  If this is the
cause of nuclear activity in quasars found in sub-{\l*} galaxies,
then we expect that the frequency of AGNs should be higher
in clusters at $z \sim 0.3-0.5$ that show the Butcher-Oemler effect.

We find the data to check this prediction to be extremely puzzling.
When they first embarked on their decade long spectroscopic study
of these clusters, Dressler and Gunn (1982, 1983, see also Dressler 1987) 
reported
that the frequency of AGNs in these clusters was much higher than the
1\% found in $z=0$ clusters (Dressler, Thompson and Shectman 1985).
A high fraction of AGNs was confirmed by Lavery and Henry (1989), but
Couch and Sharpless (1985) found that only 1 in 112 of the blue galaxies
they examined were AGNs.  

We find no clear consensus on this issue and this remains a testable prediction of the
model.  We do note that our first and third predictions have some
redundancy.   Quasars at intermediate redshifts could not lie
in clusters rich enough to be classified by Abell 
(Yates, Miller and Peacock 1989,
and Yee and Ellingson 1993) if nuclear activity is not enhanced
in clusters.

\subsection{Are there black holes in nucleated spheroidals?}

With sensible assumptions about their duty cycle, the dead remnants
of quasars are inferred to have number densities comparable to
galaxies brighter than \l* (c.f. Haehelt and Rees 1993 and references
therein).  This has stimulated a search for massive objects in the
centers of bright galaxies (c.f.  Kormendy and Richstone 1996).
Our candidate hosts have a much greater spatial frequency
and are associated with quasars at lower redshifts
where they are relatively rare.  Hence, a tiny fraction of
the spheroidals are expected to have massive nuclei.  
However, the low velocity dispersions
of the spheroidal galaxies should make it easier to detect dark
matter in their cores. 
Peterson and Caldwell (1993)
and Nieto \etal (1990) have conducted the largest studies
of spheroidals and dwarf ellipticals.   
All together, there are fewer than two
dozen galaxies.  Of these, NGC 4486B 
(Nieto \etal 1990) is one of the most unusual.
It has an absolute magnitude, $M_B = -17.4$, but its central velocity 
dispersion is measured to be $194 \pm 6$, falling to $170 \pm 7$
when averaged over its half light radius.  It's mass-to-light ratio,
is on the high end, $M/L \sim 7$, but hardly extreme.
The detection of low level nuclear activity in spheroidals 
would also be of great interest.

\section{Conclusion}

Recent HST results have challenged the conventional picture that quasars
are associated with high luminosity hosts.  We have presented a model---
galaxy harassment---that rapidly channels gas into 
the centers of low luminosity hosts.
This model has several predictions.  We find evidence that most if not
all of the quasars with sub-{\l*} hosts are in the high density environments
that lead to harassment.   The HST images of host candidates by BKS95
shows tantalizing evidence of the distortions associated with harassment.
In attempting to test the prediction of 
a higher frequency of AGNs in Butcher-Oemler 
clusters, we find some support amidst  active controversy.

\acknowledgments

This research was supported by  NASA's
High Performance Computing
and Communications Earth and Space Sciences Program,
NASA's Astrophysical Theory Program and NASA's Long Term
Space Astrophysics Program.  Our search for environmental
information was aided by the Nasa Extragalactic Database 
(NED)\footnote{The NASA/IPAC Extragalactic Database (NED) is operated by              
the Jet Propulsion Laboratory, California Institute of Technology,         
under contract with the National Aeronautics and Space Administration.}.

%

\clearpage
\centerline{\bf Figure Captions}
\bigskip


\noindent{\bf Figure 1} Snapshots of the evolution of the gas disk
in one of the harassed galaxies that we simulated.  The full time
evolution can be seen in the last segment of the video.
The disk shown is the one where 90\% of the gas is transported
into the central kpc (see Figure 2). 

\bigskip

\noindent{\bf Figure 2}  The time 
evolution of the fraction of gas
that lies within 1 kpc of the center of the galaxy.  On average,
more than half of the gas lies within this radius within a few 
Gyr---in one of the three cases, 90\% of the gas 
is in the center.  In the two cases where the most mass
is pushed into the center, about half of it 
is transferred during an interval of just 100-200 Myr.

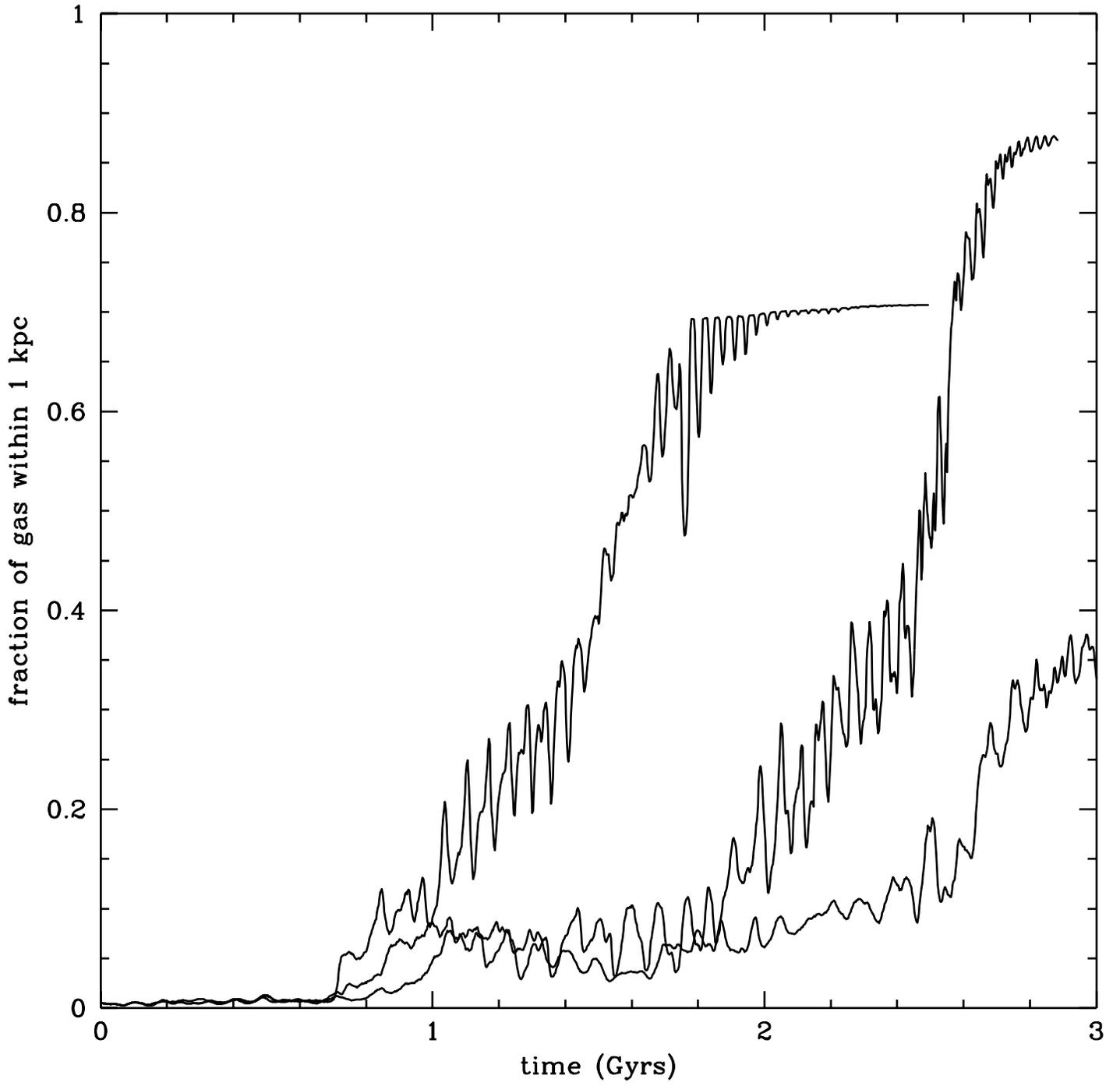
\end{document}